\documentstyle[12pt]{article}
\newcommand{\bea}{\begin{eqnarray}}
\newcommand{\eea}{\end{eqnarray}}
\newcommand{\be}{\begin{equation}}
\newcommand{\ee}{\end{equation}}
\newcommand{\p}{\prime}
\newcommand{\nn}{\nonumber}
\newcommand{\rf}[1]{(\ref{#1})}

\begin{document}
\begin{center}

{\bf DYNAMICAL SYSTEMS WITH FIRST- AND SECOND-CLASS
CONSTRAINTS AND THE SECOND NOETHER THEOREM}\\
\bigskip
{\large S.A.Gogilidze$^1$, Yu.S.Surovtsev$^2$} \\
$^1${\it Tbilisi State University, Tbilisi, University St.9,
380086 Republic of Georgia}, \\

$^2${\it Bogoliubov Laboratory of Theoretical Physics, Joint Institute for Nuclear
Research,\\ 141 980 Dubna, Moscow Region, Russia}
\end{center}
\begin{abstract}
Dynamical systems, described by Lagrangians with first- and second-class
constraints, are investigated. In the Dirac approach to the generalized
Hamiltonian formalism, the classification and separation of the first- and
second-class constraints are presented with the help of passing to an
equivalent canonical set of constraints. The general structure of
second-class constraints is clarified. The method of constructing
the generator of symmetry transformations in the second Noether theorem is
given. It is proved that second-class constraints do not contribute to the
transformation law of the local symmetry which entirely is entirely stipulated
by all the first-class constraints.
\end{abstract}
\section{Introduction}
The role of the second Noether theorem as a basis of costrained theories is
well known. Our objective is to construct a generator of local-symmetry
transformations in this theorem for the general case of dynamical systems with
first- and second-class constraints. A number of works is devoted to this
matter beginning with the basic works by Dirac \cite{Dirac}. For systems with
only first-class constraints this problem was solved in our previous papers
\cite{GSST}-\cite{GSST:tmf2}. However, there have recently appeared papers
\cite{Sugano-Kimura,Lusanna} in which one even asserts that second-class
constraints contribute also to a generator of gauge transformations which
become global in the absence of first-class constraints \cite{Sugano-Kimura}.
In the present work, we extend our scheme to theories also with second-class
constraints and prove that these constraints do not contribute to the
local-symmetry transformation law. For this aim, in the framework of the
generalised Hamiltonian formalism (with no modifications) we first outline a
separation scheme of constraints into the first- and second-class ones which
is based on passing to an equivalent canonical set of constraints. Just the
latter will be needed in what follows when deriving the local-symmetry
transformations. Note that only in recent years there have appeared real
schemes of such separation of constraints \cite{Batlle-GPR,Chaich-Mart}
proposed, however, for a modified generalised Hamiltonian formalism. Enough
complete exposition of the structure of theories with second-class constraints
is given in the works \cite{Gitman,Pavlov}.

\section{Separation of First- and Second-Class Constraints}
Let us consider a system with a number of degrees of freedom $N$ described
by a degenerate Lagrangian $L(q,\dot q)$, where $q=(q_1,\cdots,q_N)$ and
$\dot q=dq/dt=(\dot q_1,\cdots,\dot q_N)$ are generalized coordinates and
velocities, respectively. After passing to the Hamiltonian formalism, let $A$
primary constraints be obtained in the phase space $(q,p)$. Further, in
accordance with the Dirac procedure of breeding constraints \cite{Dirac}, let
us have a complete and irreducible system of constraints $\phi_\alpha^
{m_\alpha}$, where $\alpha=1,\cdots,A$ and $m_\alpha=1,2,\cdots,M_\alpha~~
(\sum_{\alpha=1}^ {A}M_\alpha=M).$ Let
\be \label{rank-2R} \mbox{rank}
\left\|\{\phi_\alpha^{m_\alpha},\phi_\beta^{m_\beta}\}\right\| =2R<M,
\ee
which implies the presence of $2R$ constraints of second class $\Psi_a^{m_a}$
and $M-2R$ constraints of first class $\Phi_\alpha^{m_\alpha}$ \cite{Dirac}.
The constraint sets $(\Phi,\Psi)$ and $\phi_\alpha^{m_\alpha}$ are related
with each other by the equivalence transformation.
A possibility of constructing the set $(\Phi,\Psi)$ was indicated by Dirac.
However, for practical aims (for example, to elucidate the role of
second-class constraints in gauge transformations \cite{GSST}) the knowledge
of an explicit form of the set $(\Phi,\Psi)$ is needed. Pass to the
$(\Phi,\Psi)$ set through several successive stages.

So, let us consider the antisymmetric matrix ${\bf K}^{11}$ with elements
$K_{\alpha\beta}^{11}=\{\phi_\alpha^1,\phi_\beta^1\}$, and let
\be \label{rank-A_1}
\mbox{rank} \left\|K_{\alpha\beta}^{11}\right\|~\stackrel{\Sigma_1}{=}A_1=2R_1<A
\ee
($\stackrel{\Sigma_1}{=}$ means this equality to hold on the primary
constraint surface $\Sigma_1$), i.e. $A_1$ primary constraints exhibit their
nature of second class already at this stage provided that subsequent stages
of our procedure do not change the established properties. One can regard the
principal minor of rank $A_1$, disposed in the left upper corner of the matrix
${\bf K}^{11}$, to be nonzero. Write down
\be \label{PB-phi^1-phi^1}
\{\phi_\alpha^1,\phi_\beta^1\}=f_{\alpha\beta\gamma}~\phi_\gamma^1+
D_{\alpha\beta},\quad \alpha,\beta,\gamma=1,\cdots,A,
\ee
where ~~$D_{\alpha\beta}\stackrel{\Sigma_1}{=}F_{\alpha\beta}.$~
Among $F_{1\alpha}~(\alpha=2,\cdots,A_1)$ at least one element is nonzero in
accordance with the supposition \rf{rank-A_1}. Let $F_{12}\neq 0$.
Pass to a new set of constraints:
\be \label{phi-to-new1-phi}
\;^1\phi_1^1\;=\;\phi_1^1\;,\quad \;^1\phi_2^1\;=\;\phi_2^1\;,\quad
\;^1\phi_\alpha^1\;=\;\phi_\alpha^1\;+\;^1u_{\alpha 1}\;\phi_1^1~+\;^1u_
{\alpha 2}\;\phi_2^1~,\quad~~\alpha=3,\cdots,A.
\ee
The left superscripts indicate a stage of our procedure and will be omitted
in the resultant expressions. Determine
~$\;^1u_{\alpha 1}\;=\;D_{2\alpha}/D_{12}\;,\quad \;^1u_{\alpha 2}\;=-D_
{1\alpha}/D_{12}~,
\quad ~\alpha=3,4,\cdots,A,$\\
to guarantee the fulfilment of the requirements:
~$\;\{^1\phi_1^1\;,\;^1\phi_\alpha^1\}~\stackrel{\Sigma_1}{=}\;0,\quad
\;\{^1\phi_2^1\;,\;^1\phi_\alpha^1\}~\stackrel{\Sigma_1}{=}\;0,~$ \\
and to obtain:
~$^1D_{12}=\;-^1D_{21}\;\stackrel{\Sigma_1}{=}\;^1F_{12}\;=F_{12}~\neq 0\;,
\quad ^1D_{\alpha\beta}~\stackrel{\Sigma_1}{=}F_{\alpha\beta}~=0,\quad \alpha=
1,2,~~\beta=3,4,\cdots,A. $\\
So, by means of the transformation
~$\;^1\phi_\alpha^1=\;^1\Lambda_{\alpha\beta}\;\phi_\beta^1,\quad
\mbox{det}\|^1\Lambda_{\alpha\beta}\|=1,$~
we obtain
\be \label{^1K11}
^1K_{\alpha\beta}^{11}~=\;\{^1\phi_\alpha^1\;,\;^1\phi_\beta^1\}\;=
\;^1\Lambda_{\alpha\sigma}\;^1\Lambda_{\beta\tau}\;K_{\sigma\tau}^{11}\;+~
O(\phi_\alpha^1),
\ee
$$\;^1{\bf K}^{11}\;\stackrel{\Sigma_1}{=}\left( \begin{array}{cc}
F_{12}\cdot {\bf J} & {\bf O} \\
{\bf O} & \|^1{\cal F}_{\alpha\beta}\|(\alpha,\beta=3,4,\cdots,A)
\end{array}  \right), $$
where
${\bf J}=\left( \begin{array}{cc}
0 & 1 \\
-1 & 0
\end{array}  \right)$ and ${\bf O}$ are zero blocks, and  $\|^1{\cal F}_
{\alpha\beta}\|$ is $(A-2)\times (A-2)$-block which must be reduced to the
quasidiagonal form at the next stages of the procedure. This procedure must be
iterated $R_1=A_1/2$ times, and we shall obtain
\bea \label{^R_1K11}
\;^{R_1}{\bf K}^{11}\;\stackrel{\Sigma_1}{=}\left( \begin{array}{c|c}
{\begin{array}{cccc}
F_{12}\cdot {\bf J} & {\bf O} & \ldots & {\bf O}\\
{\bf O} & ^1F_{34}\cdot {\bf J} & \ldots & {\bf O}\\
\vdots & \vdots & \ddots & \vdots \\
{\bf O} & {\bf O} & \ldots & \;^{R_1-1}F_{A_1-1~A_1}\cdot {\bf J}\;
\end{array}} & {\bf O}\\
\hline
{\bf O} & {\bf O}
\end{array}  \right).
\eea
The corresponding equivalent set of primary constraints is determined by the
relation:
$$\;^{R_1}\phi_\alpha^1~=\;^{R_1}\Lambda_
{\alpha\beta}\;{^{R_1-1}\Lambda_{\beta\gamma}}\cdots {^1\Lambda_{\sigma\tau}}~
{^1\phi_\tau^1}~=\overline{\Lambda}_{\alpha\beta}\;\phi_\beta^1~,\quad
\mbox{det}\overline{{\bf \Lambda}}=1.$$
Denoting the second-class constraints by the letter $\psi(\Psi)$, we now have
the following set of primary constraints:  ~~$
[\psi_{a_1}^1]_{a_1=1}^{A_1}~,\quad [\phi_{\alpha_1}^1]_{\alpha_1=1}^{A-A_1}$~
with the properties
\be \label{PB-psi1-psi1}
\{\psi_{a_1}^1,\psi_{b_1}^1\}\stackrel{\Sigma_1}{=}\left\{\begin{array}{ll}
F_{a_1b_1}\neq 0,\quad& a_1=2k+1,~b_1=2k+2~(k=0,1,\cdots,A_1-2) \\
{}& \mbox{and conversely},\\
0& \mbox{in other cases},
\end{array} \right.
\ee
\be \label{PB-psi1-phi1}
\{\psi_{a_1}^1,\phi_{\alpha_1}^1\}~\stackrel{\Sigma_1}{=}0,\qquad
\{\phi_{\alpha_1}^1,\phi_{\beta_1}^1\}~\stackrel{\Sigma_1}{=}0.
\ee
It is clear that the constraints $\psi_{a_1}^1$ do not generate secondary
constraints. Furthermore, with the help of the transformation
$~\;^1\phi_{\alpha}^{m_\alpha}~=\phi_{\alpha}^{m_\alpha}~+~
C_{\alpha b_1}^ {m_\alpha}~\psi_{b_1}^1~$\\ one can attain that
~~~~~~~~~~~~~~$\{\psi_{a_1}^1,\phi_{\alpha}^{m_\alpha}\}~\stackrel{\Sigma_1}{=}0,\quad
m_\alpha=2,\cdots,M_\alpha.$

Now let us turn to $\phi_{\alpha_1}^{m_{\alpha_1}},~\alpha_1=1,\cdots,A-A_1.$
Let
~$\mbox{rank}\left\|\{\phi_{\alpha_1}^1,\phi_{\beta_1}^2\}\right\|~
\stackrel{\Sigma}{=}A_2<A-A_1$\\
($\Sigma$ is the surface of all constraints).
Further, one has to proceed by analogy with the proceding case with the
differences that now the transformation of primary constraints will generate
the transformation of secondary constraints according to Dirac's scheme, and
in this case we have
$$\{\phi_{\alpha_1}^1,\phi_{\beta_1}^2\}~\stackrel{\Sigma_2}{=}
~\{\phi_{\beta_1}^1,\phi_{\alpha_1}^2\}.$$
Therefore, ${\bf K}^{12}|_{\Sigma}$ will be given in the diagonal form with
only nonvanishing diagonal elements.
So, due to
~$\{\psi_{a_2}^2,\psi_{b_2}^2\}~\stackrel{\Sigma}{=}~0~(a_2,b_2=1,\cdots,A_2),
$~ we obtain, at this stage, $A_2$ one-linked chains of second-class
constraints which are in involution on $\Sigma$ with each other and with all
other constraints.

Next one must consider the constraints $\phi_{\alpha_2}^{m_{\alpha_2}},~~
\alpha_2= 1,\cdots,A-A_1-A_2$. Let\\
\hspace*{4cm}$\mbox{rank} \left\|\{\phi_{\alpha_2}^1,\phi_{\beta_2}^3\}\right\|~
\stackrel{\Sigma_3}{=}A_3=2R_3<A-A_1-A_2.$\\
With the help of the Jacobi identity we obtain
~~$\{\phi_{\alpha_2}^1,\phi_{\beta_2}^3\}~\stackrel{\Sigma_3}{=}
~-\{\phi_{\beta_2}^1,\phi_{\alpha_2}^3\}. ~$ \\
Proceeding by analogy with the previous case, we obtain the matrix
${\bf K}^ {13}|_{\Sigma}$  in the quasidiagonal form with only nonvanishing
elements along the principal diagonal $F_{a_3b_3}^{1~3}\neq 0$~
(where if $a_3=2k+1,~b_3= 2k+2$ and conversely; $k=0,1,\cdots,A_3-2$).

Now we have the relations
$$\{\psi_{a_3}^2,\psi_{b_3}^3\}~\stackrel{\Sigma}{=}
\{\psi_{a_3}^3,\psi_{b_3}^3\}~\stackrel{\Sigma}{=}~0,\quad
\{\psi_{a_3}^2,\psi_{b_3}^2\}~\stackrel{\Sigma_3}{=}-\{\psi_{a_3}^1,
\psi_{b_3}^3\}, \quad a_3,b_3=1,\cdots,A_3. $$
Thus, two-linked doubled chains of second-class constraints are obtained.
Constraints of such different formations are in involution on $\Sigma$ with
each other and with all other constraints since all the previously
established properties are kept.

This procedure must be continued. We emphasize that every subsequent stage
preserves the properties of transformed constraints, which were obtained at
the preceding stages, and, therefore, the secondary, tertiary, etc.
constraints do not mix themselves into primary constraints. If after carrying
out certain $n$-th stage it is found that
$$\mbox{rank}\left\|\{\phi_{\alpha_n}^{m_{\alpha_n}},\phi_{\beta_n}^{m_
{\beta_n}}\}\right\|~ \stackrel{\Sigma}{=}~0,\qquad
\alpha_n,\beta_n=1,\cdots,A-\sum\nolimits_{1}^{n}A_j,$$
these remaining constraints $\phi_{\alpha_n}^{m_{\alpha_n}}$ are all of
first class.

So, the final set of constraints $(\Phi,\Psi)$ is obtained from the initial
one $\phi_\alpha^{m_\alpha}$ by the equivalence transformation.

\section{Local Symmetry Transformations}

A group of phase-space coordinate transformations, which map each
solution of the Hamiltonian equations of motion into the solution of the same
equations, will be called a symmetry transformation. Under these
transformations the action functional is quasi-invariant within a surface
term. So, we shall consider the action
\be \label{S}
S=\int_{t_1}^{t_2}dt~(p~\dot q - H_T)
\ee
where
\be \label{H_T}
H_T = H^\p + u_\alpha\Phi_\alpha^1,\qquad
H^\p = H_c+\sum\nolimits_{i=1}^{n}({\bf K}^{1~i})_{b_i~a_i}^{-1}\{\Psi_{a_i}^i,H_c\}
\Psi_{b_i}^1,
\ee
$H_c$ is the canonical Hamiltonian, $u_\alpha$ are the Lagrange multipliers.
We shall require a quasi-invariance of action \rf{S} with respect to
transformations:
\bea \label{q-q^prime}
\left\{\begin{array}{ll} q_i^\p = q_i+\delta q_i ,& \quad
\delta q_i = \{q_i , G\} ,\\
p_i^\p = p_i+\delta p_i ,& \quad\delta p_i =\{p_i , G\}. \end{array}\right.
\eea
The generator $G$ will be looked for in the form
\be \label{G-Dirac}
G=\varepsilon_\alpha^{m_\alpha}\Phi_\alpha^{m_\alpha} +
\eta_{a_i}^{m_{a_i}}\Psi_{a_i}^{m_{a_i}}.
\ee

So, under transformations \rf{q-q^prime} we have
\be \label{delta-S}
\delta S=\int_{t_1}^{t_2}dt\bigl[\delta p~\dot q+p~\delta \dot q-
\delta H_T \bigr]= \int_{t_1}^{t_2}dt\bigl[\frac{d}{dt}(p~\frac{\partial G}
{\partial p}-G)-\frac{\partial G}{\partial t}-\{G,H_T\}\bigr].
\ee
>From \rf{delta-S} we see: in order that transformations \rf{q-q^prime} were
the symmetry ones, it is necessary that
\be \label{sym-condition}
\frac{\partial G}{\partial t}+\{G,H_T\}~\stackrel{\Sigma_1}{=}0.
\ee
The fact that the last equality must be realized on the primary-constraint
surface $\Sigma_1$, can easily be interpreted if one remembers that $\Sigma_1$
is the whole $(q,\dot q)$-space image in the phase space. Since under the
operation of the local-symmetry transformation group the configuration space
is being mapped into itself in a one-to-one manner, the one-to-one mapping of
$\Sigma_1$ into itself corresponds to this in the phase space. Therefore, at
looking for the generator $G$, it is natural to require also $\Sigma_1$ to be
conserved under transformations \rf{q-q^prime}, i.e. the requirement
\rf{sym-condition} must be supplemented by the demands \be \label{PB-G-Psi1}
\bigl\{G,\Psi_{a_i}^1\bigr\}~\stackrel{\Sigma_1}{=}0,  \qquad
\bigl\{G,\Phi_\alpha^1\bigr\}~\stackrel{\Sigma_1}{=}0.
\ee
Furhter, we shall use the following Poisson brackets:
\bea
&& \bigl\{\Phi_\alpha^{m_\alpha},H^\p\bigr\} =
g_{\alpha~~\beta}^{m_\alpha m_\beta}~\Phi_\beta^{m_\beta},\quad
m_\beta=1,\cdots,m_\alpha+1, \label{PB-Phi-H^prime}\\
&& \bigl\{\Psi_{a_i}^{m_{a_i}},H^\p\bigr\} = \bar{g}_{{a_i}~~\alpha}^{m_{a_i}
m_\alpha}~\Phi_\alpha^{m_\alpha}+\sum\nolimits_{k=1}^{n}h_{{a_i}~~{b_k}}^{m_{a_i}
m_{b_k}}~\Psi_{b_k}^{m_{b_k}},\quad m_{b_n}=m_{a_i}+1,\label{PB-Psi-H^prime}\\
&& \bigl\{\Phi_\alpha^{m_\alpha},\Phi_\beta^{m_\beta}\bigr\} =
f_{\alpha~~\beta~~\gamma}^{m_\alpha m_\beta m_\gamma}~\Phi_\gamma^{m_\gamma}
\label{PB-Phi-Phi}.
\eea
The general properties of the structure functions can be extracted from the
consideration of the previous section.

So, from eqs.\rf{sym-condition} and \rf{G-Dirac} with taking account of
\rf{PB-Phi-H^prime}-\rf{PB-Phi-Phi} we write down
\bea \label{eps.eta-Phi-Psi}
&&\Bigl (\dot \varepsilon_\alpha^{m_\alpha}+\varepsilon_\beta^{m_\beta}
g_{\beta~~\alpha}^{m_\beta m_\alpha}+\sum\nolimits_{i=1}^{n}\eta_{a_i}^{m_{a_i}}
\bar{g}_{{a_i}~~\alpha}^{m_{a_i} m_\alpha}~\Bigr)\Phi_\alpha^{m_\alpha}\nn\\
&&
+\sum\nolimits_{i=1}^{n}\Bigl(\dot{\eta}_{a_i}^{m_{a_i}}+\sum\nolimits_{k=1}^{n}\eta_{b_k}^
{m_{b_k}} h_{{b_k}~~{a_i}}^{m_{b_k} m_{a_i}}~\Bigr)\Psi_{a_i}^{m_{a_i}}
+u_\alpha\bigl\{G,\Phi_\alpha^1\bigr\} ~\stackrel{\Sigma_1}{=}0.
\eea
Taking into consideration \rf{PB-G-Psi1}, we have~~
$u_\alpha\bigl\{G,\Phi_\alpha^1\bigr\} ~\stackrel{\Sigma_1}{=}0.$\\
Then, in view of the functional independence of constraints, in order to carry
out the equality \rf{eps.eta-Phi-Psi}, one must demand the coefficients of the
constraints  $\Phi_\alpha^{m_\alpha}$ and $\Psi_{a_i}^{m_{a_i}}$ to vanish.

Before analyzing these conditions to satisfy the equality \rf{eps.eta-Phi-Psi},
let us consider in detail the conditions of the $\Sigma_1$ conservation
\rf{PB-G-Psi1} starting from the former. Its realization would mean
$$\sum\nolimits_{k=1}^{n}\eta_{b_k}^{m_{b_k}}\bigl\{\Psi_{b_k}^{m_{b_k}},\Psi_{a_i}^1
\bigr\}~\stackrel{\Sigma_1}{=}0.$$
In this equality for each value of $a_i$ in the double sum over $k$ and over
$b_k$ the only nonvanishing Poisson brackets are those at $b_k=a_i,~M_{b_k}=
i$; therefore
\be \label{eta-i=0}
\eta_{a_i}^i=0\quad
\mbox{for}~ i=1,\cdots,n.
\ee
Now considering the requirement for the coefficients of the
constraints $\Psi_{a_i}^{m_{a_i}}$ in eq.\rf{eps.eta-Phi-Psi} to vanish
\be \label{eq:eta}
\dot{\eta}_{a_i}^{m_{a_i}}+\sum\nolimits_{k=1}^{n}\eta_{b_k}^{m_{b_k}}
h_{{b_k}~~{a_i}} ^{m_{b_k} m_{a_i}}~\stackrel{\Sigma_1}{=}0 \ee
we obtain with taking account of \rf{eta-i=0} that this system of algebraic
equations for unknowns $\eta_{b_k}^{m_{b_k}}$ has only a trivial solution
$\eta_{b_k}^ {m_{b_k}}=0$ since
$$\mbox{det}\|h_{b_i~~~~~a_i}^{i-k-1~i-k}\|\neq
0\quad(~\mbox{and}~~
\mbox{det}\|g_{\beta~~~~~~~\alpha}^{M_\alpha-k-1~M_\alpha-k}\|\neq 0~),$$
that is easily proved as a consequence of functional independence of all
constraints with the help of the method by contradiction \cite{GSST:tmf1}.

So, it is proved that the second-class constraints do not contribute to a
generator of local-symmetry transformations.

Returning to the second condition of the $\Sigma_1$ conservation \rf{PB-G-Psi1}
under transformations \rf{q-q^prime}, we see that it will be fulfilled if
\be \label{ideal}
\bigl\{\Phi_\alpha^1,\Phi_\beta^{m_\beta}\bigr\}=f_{\alpha~~\beta~~\gamma}^
{1~~ m_\beta~ 1}~\Phi_\gamma^1 .
\ee
This relation means a quasi-algebra of the special form where first-class
primary constraints make an ideal of quasi-algebra formed by all first-class
constraints.

To determine the multipliers $\varepsilon_\alpha^{m_\alpha}$ in the generator
\rf{G-Dirac}, we have only a requirement of vanishing coefficients of
constraints $\Phi_\alpha^{m_\alpha}$ in \rf{eps.eta-Phi-Psi} \cite{GSST}:
\be \label{eq:eps}
\dot \varepsilon_\alpha^{m_\alpha}+\varepsilon_\beta^{m_\beta}
g_{\beta~~\alpha}^{m_\beta m_\alpha}=0,\qquad m_\beta= m_\alpha-1,\cdots,
M_\alpha .
\ee
In system of equations \rf{eq:eps} the number of unknowns exceeds the number
of equations by the number $F=A-\sum_{i=1}^{n}A_i$ of the first-class primary
constraints; therefore, the system \rf{eq:eps} can be solved to within $F$
arbitrary functions.
Introducing arbitrary functions ~~$\varepsilon_\alpha\equiv\varepsilon_\alpha^
{M_\alpha}~(\alpha = 1,\cdots, F)$~ and inserting them into equations
\rf{eq:eps}, we obtain a system of algebraic equations, that always have a
solution \cite{GSST}.
Then, the generator of local-symmetry transformations takes the form:
\be \label{G}
G=B_{\alpha~~\beta}^{m_\alpha m_\beta}\phi_\alpha^{m_\alpha}\varepsilon_\beta^
{(M_\alpha-m_\beta)} , \qquad m_\beta=m_\alpha,\cdots,M_\alpha,
\ee
where
~~$\varepsilon_\beta^{(M_\alpha-m_\beta)}\equiv ({{d^{M_\alpha-m_\beta}}/
{dt^{M_\alpha-m_\beta}}})\varepsilon_\beta (t),$~
and $B_{\alpha~~\beta}^{m_\alpha m_\beta}$ are, generally speaking, functions
of $q$ and $p$ and their derivatives up to the order $M_\alpha-m_\alpha-1$.
The obtained generator \rf{G} satisfies the group property
~~$\{G_1,G_2\}=G_3$~
where the transformation $G_3$ is realized by carrying out two successive
transformations $G_1$ and $G_2$.
The amount of group parameters $\varepsilon_\alpha (t)$, which determine a
rank of a quasigroup of these transformations, equals the number of primary
constraints of first class. As can be seen from formula \rf{G}, the
transformation law may include both arbitrary functions $\varepsilon_\alpha(t)$
and their derivatives up to and including the order $M_\alpha-1$; the highest
derivatives $\varepsilon_\alpha^{(M_\alpha-1)}$ should always be present.

Several remarks: ~(i) Though the condition \rf{ideal} means a constraint
quasi-algebra of a special form, it is fulfilled in most of the physically
interesting theories. However, in the available literature there are examples
of Lagrangians where this condition \rf{ideal} does not hold, e.g., Polyakov's
string and other model Lagrangians \cite{GSST:tmf2}. In the case of theories
with only first-class constraints we have shown \cite{GSST:tmf2} that there
always exist equivalent sets of constraints, for which the condition
\rf{ideal} holds valid, and given a method of passing to a set like these.
This proof and method are valid also for theories with first- and second-class
constraints. ~(ii) The corresponding transformations of local symmetry in the
Lagrangian formalism are determined in the following way:  \be \delta q_i(t)
=\{q_i(t), G\}\bigr|_{p=\frac{\partial L}{\partial \dot q}},\qquad \delta\dot
q(t) = \frac{d}{dt}\delta q(t).  \ee (iii) If time derivatives of $q$ and $p$
are present in $B_{\alpha~~\beta}^{m_\alpha m_\beta}$, the Poisson brackets in
\rf{q-q^prime} are not determined. This problem was solved in our previous
work \cite{GSST:tmf1}. It is shown that local-symmetry transformations are
canonical in the extended (by Ostrogradsky) phase space using the formalism
of theories with higher derivatives \cite{Ostr,Nesterenko}.\\

{\Large Examples:}
~1. Consider the Lagrangian \cite{Sugano-Kimura}
\be \label{exam:L}
L=(\dot q_1 + \dot q_2) q_3+\frac{1}{2}{\dot q_3}^2-\frac{1}{2}{q_2}^2.
\ee
In the phase space we obtain two primary  constraints
~~$\phi_1^1=p_1-q_3, \quad \phi_2^1=p_2-q_3$~~ and the total Hamiltonian:
\be \label{exam:H_T}
H_T=\frac{1}{2}({p_3}^2 + {q_2}^2) +u_1\phi_1^1+u_2\phi_2^1.
\ee
>From the self-consistency conditions of the theory we derive two secondary
constraints
~~$\phi_1^2=p_3, \quad \phi_2^2=p_3-q_2$~~and the Lagrangian multipliers
~~$u_1=u_2=0.$
Let us calculate the matrix~
${\bf W}=\left\|{\bf K}^{m_\alpha m_\beta}\right\|=\left\|\{\phi_\alpha^
{m_\alpha},\phi_\beta^{m_\beta}\}\right\|$: \\
\hspace*{5.8cm}${\bf W}=\left(\begin{array}{rrrr}
0 & ~~0 & -1 & -1 \\
0 & ~~0 & -1 & -2 \\
1 & ~~1 & 0 & 0 \\
1 & ~~2 & 0 & 0
\end{array}  \right).$\\
We see that ~$\mbox{rank}{\bf W}=4$, i.e. all constraints are of second class;
therefore, ${\bf W}$ have a quasidiagonal (antisymmetric) form. Performing
our procedure, we pass to the equivalent canonical set of constraints
$\Psi_a^ {m_a}$:\\
\hspace*{3.5cm}$\left(\begin{array}{c} \Psi_1^1\\ \Psi_2^1\\ \Psi_1^2\\
\Psi_2^2 \end{array} \right)= \left(\begin{array}{rrrr} 1 & 0 & ~~0 & 0 \\ 1 &
-1 & ~~0 & 0 \\ 0 & 0 & ~~1 & 0 \\ 0 & 0 & ~~1 & -1 \end{array} \right)
\left(\begin{array}{c} \phi_1^1\\ \phi_2^1\\ \phi_1^2\\ \phi_2^2 \end{array}
\right)= \left(\begin{array}{c} p_1-q_3\\ p_1-p_2\\ p_3\\ -q_2 \end{array}
\right).$\\
For the last set of constraints the quasidiagonal form of
${\bf W}$  will have a canonical structure \\
\hspace*{5.3cm}${\bf W}^\p=\left(\begin{array}{rrrr}
0 & ~~0 & -1 & 0 \\
0 & ~~0 & 0 & -1 \\
1 & ~~0 & 0 & 0 \\
0 & ~~1 & 0 & 0
\end{array} \right).$\\
For quasi-invariance of the action with respect to transformations
\rf{q-q^prime} with the generator
$$G=\eta_1^1~\Psi_1^1+\eta_2^1~\Psi_2^1+\eta_1^2~\Psi_1^2
+\eta_2^2~\Psi_2^2,$$
it is necessary to realize the condition \rf{PB-G-Psi1} of the $\Sigma_1$
conservation under these transformations:\\
\hspace*{5.2cm}$\bigl\{G,\Psi_a^1\bigr\}~\stackrel{\Sigma_1}{=}0,\qquad
a=1,2.$\\ From here we obtain ~$\eta_1^2=\eta_2^2=0.$  Next from \rf{eq:eta}
we establish ~$\eta_1^1=\eta_2^1=0,$  i.e. the second-class constraints of the
system $\Psi_a^ {m_a}$ do not generate transformations of either local
symmetry or global one.

2. Now consider the well-known case of spinor electrodynamics. In the phase
space ($A_\mu,\psi,\overline{\psi}$~ and~ $\pi_\mu,p_\psi,p_{\overline{\psi}}$
~are the generalized coordinates and momenta, respectively; ~$F_{\mu\nu}=
\partial_\mu A_\nu-\partial_\nu A_\mu$) we have the canonical Hamiltonian
\be \label{el.dyn:H_T}
H_c= \int d^3x \Bigl[\frac{1}{4} F_{ij}F^{ij}+\frac{1}{2}\pi^i \pi^i+
\pi_i\partial_i A_0+ iep_\psi A_0 \psi +i\overline{\psi}\gamma_i(\partial_i
-ieA_i)\psi+m\overline{\psi} \psi \Bigr].
\ee
and three primary constraints:
~$\phi_1^1=\pi_0, \quad \phi_2^1=p_\psi-i\overline{\psi}\gamma_0, \quad
\phi_3^1=p_{\overline{\psi}}.$~
Among the conditions of the constraint conservation in time~ $\dot{\psi}_i^1=
0~(i=1,2,3)$~ two last ones serve for determining the Lagrangian multipliers.
>From the first condition we obtain one secondary constraint
~$\phi_1^2= \partial_i\pi^i-iep_\psi \psi.$~
Calculating the matrix ~${\bf W}=
\left\|\{\phi_\alpha^{m_\alpha},\phi_\beta^{m_\beta}\}\right\|$:\\
\hspace*{3.6cm}${\bf W}=\delta(x-x^\p)\left(\begin{array}{cccc}
0 & ~~0 & 0 & 0 \\
0 & ~~0 & i\gamma_0 & -iep_\psi \\
0 & ~~i\gamma_0 & 0 & -e\gamma_0 \psi \\
0 & ~~iep_\psi & e\gamma_0 \psi & 0
\end{array}  \right),$ \\
we see that ~$\mbox{rank}{\bf W}=2$; therefore, two constraints are of second
class and the other two are of first class. Now implementing our procedure, we
shall pass to the canonical set of constraints by the equivalence
transformation \\
\hspace*{1.3cm}$\left(\begin{array}{c}
\Psi_1^1\\ \Psi_2^1\\ \Phi_1^1\\ \Phi_1^2 \end{array}  \right)=
\left(\begin{array}{cccc}
1 & 0 & ~0 & ~~~0 \\
0 & 1 & ~0 & ~~~0 \\
0 & 0 & ~1 & ~~~0 \\
ie\psi & -ie\overline{\psi} & ~0 & ~~~1
\end{array}  \right)
\left(\begin{array}{c}
\phi_2^1\\ \phi_3^1\\ \phi_1^1\\ \phi_1^2 \end{array}  \right)=
\left(\begin{array}{c}
p_\psi-i\overline{\psi}\gamma_0\\ p_{\overline{\psi}}\\ \pi_0\\
\partial_i\pi^i-ie(p_\psi \psi+\overline{\psi}p_{\overline{\psi}})
\end{array}  \right)$\\
where the constraints are already separated into the ones of first and second
class since now the matrix ~${\bf W}$~ has the form\\
\hspace*{4cm}${\bf W}^\p=\delta(x-x^\p)\left(\begin{array}{cccc}
0 & i\gamma_0 & ~0 & ~~0 \\
i\gamma_0 & 0 & ~0 & ~~0 \\
0 & 0 & ~0 & ~~0 \\
0 & 0 & ~0 & ~~0
\end{array} \right).$\\
Further, we look for the generator $G$ in the form
~~~$G=\int d^3x\bigl[\eta_1^1~\Psi_1^1+\eta_2^1~\Psi_2^1+
\varepsilon_1^1~\Phi_1^1+ \varepsilon_1^2~\Phi_1^2\bigr].$~\\
>From the first condition \rf{PB-G-Psi1} of conservation of $\Sigma_1$ under
the transformations \rf{q-q^prime} we derive ~$\eta_1^1=\eta_2^1=0,$ i.e. the
constraints of second class do not contribute to $G$. The second condition
\rf{PB-G-Psi1} is realized because
~$\bigl\{\Phi_1^1,\Phi_1^2\bigr\}=0.$~~ Eq.\rf{eq:eps} accepts the form:
~~$\dot \varepsilon_1^2 -\varepsilon_1^1 = 0,$~~i.e. ~$\varepsilon_1^1=\dot
\varepsilon$~ where $\varepsilon\equiv \varepsilon_1^2$.~ Therefore, we have \\
$$G=\int d^3x\Bigl\{\dot \varepsilon \pi_0+\varepsilon[\partial_i
\pi^i-ie(p_\psi \psi+ \overline{\psi}p_{\overline{\psi}})]\Bigr\},$$
from which it is easily to obtain the gauge transformations in the phase space
and the well-known transformation rule:  ~$\delta A_\mu=\partial_\mu
\varepsilon,\quad \delta \psi=ie\varepsilon\psi, \quad \delta
\overline{\psi}=-ie\varepsilon\overline{\psi}.$

\end{document}